\documentclass[aps,prd,preprint,superscriptaddress,tightenlines,nofootinbib]{revtex4}

\usepackage{amsfonts,amsmath,amsthm,amssymb}
\usepackage{mathrsfs}
\usepackage{bm}
\usepackage{color}
\usepackage{axodraw4j}
\usepackage{pstricks}

\usepackage{epsfig}

\newcommand{\PRE}[1]{{#1}}   

\newcommand{\comment}[1]{}

\newcommand{\ci}[1]{}

\newcommand{\ke}{\rangle}
\newcommand{\br}{\langle}

\newcommand{\ba}{\begin{eqnarray}}
\newcommand{\ea}{\end{eqnarray}}
\newcommand{\be}{\begin{equation}}
\newcommand{\ee}{\end{equation}}
\newcommand{\bay}[1]{\left(\begin{array}{#1}}
\newcommand{\eay}{\end{array}\right)}

\def\xa{{\alpha}}

\def\BR{\mathbb{R}}

\def\lsq{[\![}
\def\rsq{]\!]}

\newcommand{\req}[1]{(\ref{#1})}

\newcommand{\nwc}{\newcommand}

\def\ap{\alpha'}

\nwc{\bea} {\begin{eqnarray}}
\nwc{\eea} {\end{eqnarray}}
\nwc{\nnn} {\nonumber \vspace{.2cm} \\ }

\nwc{\bda} {\bdm\ba{lcl}}
\nwc{\eda} {\ea\edm}

\nwc{\ds}  {\displaystyle}
\nwc{\ra}{\rightarrow}
\nwc{\lra}{\longrightarrow}
\def\lf{\left}\def\ri{\right}

\begin{document}

\preprint{
\hfil
\begin{minipage}[t]{3in}
\begin{flushright}
\vspace*{.4in}
MPP--2012--65\\
\end{flushright}
\end{minipage}
}

\vspace*{-5cm}
\title{Maximally Helicity Violating Disk Amplitudes, \\[2mm]
Twistors and Transcendental Integrals
\PRE{\vspace*{0.3in}} }

\author{Stephan Stieberger}
\affiliation{Max--Planck--Institut f\"ur Physik\\
 Werner--Heisenberg--Institut,
80805 M\"unchen, Germany
\PRE{\vspace*{.1in}}
}
\author{Tomasz R. Taylor}
\affiliation{Department of Physics\\
  Northeastern University, Boston, MA 02115, USA \PRE{\vspace*{.1in}}
}

\date{April 2012}

\PRE{\vspace*{3.5in}}

\begin{abstract}\vskip 3mm
  \noindent
 We obtain simple expressions for tree--level maximally helicity violating amplitudes of $N$ gauge bosons from disk world--sheets of open superstrings. The amplitudes are written in terms of \linebreak $(N-3)!$ hypergeometric integrals depending on kinematic parameters, weighted by certain kinematic factors. The integrals are transcendental in a strict sense defined in this work. The respective kinematic factors can be succinctly written in terms of ``dual'' momentum twistors.
 The amplitudes are computed by using the prescription proposed  by Berkovits and Maldacena.
\end{abstract}

\maketitle

Humans observe the world by detecting photons and other particles scattered all over the Universe. For a scientific understanding to be possible, the scattering amplitudes must contain full information not only about all physical processes, but also about the nature of spacetime. Some of the simplest amplitudes describe the scattering of gauge bosons (gluons) in (supersymmetric) Yang-Mills theories \cite{Parke:1986gb}. In 2003, Edward Witten proposed a twistor string theory description of such amplitudes \cite{Witten:2003nn}. Ever since then, there has been a growing evidence that twistors play an important role in the S-matrix description of spacetime, with the ambient spacetime possibly emerging from a twistor space.

Strings propagate in ambient spacetime. Open superstrings offer a self-consistent generalization of Yang-Mills theory and introduce the Regge slope parameter $\xa'$ that determines the mass scale $M=1/\sqrt{\xa'}$ of higher spin excitations of the gauge supermultiplet. The scattering amplitudes receive contributions from all two-dimensional world-sheets with and without boundaries. Our goal is to obtain
compact expressions for some full-fledged but possibly simplest superstring amplitudes.
We also want to see if twistors play any special role in a theory in which (spacetime) superconformal invariance is explicitly broken by the presence of mass scale $M$.
 We will consider the semi-classical contributions of disk world-sheets to multi-particle amplitudes involving gluons in maximally helicity violating (MHV) configurations which, by experience with Yang-Mills theory, are expected to have the simplest form.

Disk amplitudes have been studied for almost fifty years. Some time ago, we performed a detailed analysis of multi-gluon amplitudes, deriving explicit expressions for up to $N=7$ gluons \cite{Oprisa:2005wu,Stieberger:2006bh,Stieberger:2006te,Stieberger:2007jv,Stieberger:2007am}. More recently, more compact expression have been obtained by using the pure spinor formalism \cite{Mafra:2011nv,Mafra:2011nw}. The general structure is, however, always the same. For a specific ordering of non-abelian gauge group (Chan-Paton) factors, the $N$-gluon (partial) amplitudes are usually written as sums of $(N{-}3)!$ terms:
$\sum_{i=1}^{i=(N{-}3)!}I_i\ K_i$,
where $I_i$ are some integrals over $N{-}3$ positions $z_4,\dots,z_N$ of vertex operators at the disk boundary ($z_1,z_2,z_3$ are fixed by $PSL(2,\BR)$ invariance), while $K_i$ are certain kinematic factors that depend on the polarization vectors and momenta of external gluons. In this work, we discuss the integrals and kinematic factors describing MHV amplitudes. Instead of performing traditional computations, we follow the prescription of Berkovits and Maldacena \cite{Berkovits:2008ic} which offers a  shortcut for MHV configurations.

Berkovits and Maldacena \cite{Berkovits:2008ic} propose to compute the $N$-gluon partial disk amplitude, associated to the Chan-Paton factor Tr$(T^{a_1}\cdots T^{a_N})$, and involving two helicity minus gluons (labeled by 1 and 2) and $N{-}2$ helicity plus gluons, as the following correlation function:
\be
A_N(\lambda_k,\tilde{\lambda}_k)=\frac{\br 12\ke^4}{\br 12\ke
\br 23\ke\br 31\ke}\left\langle V_1(z_1)V_2(z_2)V_3(z_3)\int_{z_3}^{\infty}dz_4 U_4(z_4)\cdots\int_{z_{N-1}}^{\infty}dz_N U_N(z_N)\right\rangle\label{bmprop}
\ee
where
$
V_k(z_k)=e^{ip_kX(z_k)}=e^{i\lambda_k\tilde{\lambda}_kX(z_k)}
$ and
\be U_k(z_k)=(\epsilon_k^a\tilde{\lambda}_k^{\dot{a}}\partial X_{a\dot{a}}+\psi_{\dot{a}}\bar{\psi}_{\dot{b}}
\tilde{\lambda}_k^{\dot{a}}
\tilde{\lambda}_k^{\dot{b}})e^{ip_kX(z_k)}\ .\label{vert}
\ee
Here, $X$ are the standard bosonic coordinates while $\psi_{\dot{a}},~ \bar{\psi}_{\dot{b}}$ are fermions of conformal weight $(\frac{1}{2},0)$. $\epsilon^a_k$ are arbitrary ``reference spinors'' normalized by $\epsilon^a_k\lambda_{ak}=1$. In this work, we follow the same conventions as in Ref.\cite{Stieberger:2006te}, always choosing $(z_1,z_2,z_3)= (-\infty, 0, 1)$, and integrate in Eq.(\ref{bmprop}) over  $N{-}3$ real variables ordered as $z_4<\dots < z_N$ in the interval $(1,\infty)$.

It should be made clear that Eq.(\ref{bmprop}) is a conjecture supported by several tests. Berkovits and Maldacena have already checked that Eq.(\ref{bmprop}) does indeed yield correct Yang-Mills amplitudes in the $\xa'\to 0$ limit. Furthermore, they verified that for $N=4$ and $N=5$, it reproduces the known string amplitudes.
As a warm-up, we will first examine the $N=6$ case, using this as an opportunity to establish some notation.

For $N=6$, the correlator of Eq.(\ref{bmprop}) becomes
\be \left\langle V_1(z_1)V_2(z_2)V_3(z_3)
\prod_{k=4,5,6}\int dz_k(\varepsilon_k\partial X(z_k)+\psi(z_k)\tilde{\lambda}_k\bar{\psi}(z_k)\tilde{\lambda}_k)V_k(z_k)\right\rangle\label{cor6}
\ee
with the polarization vectors $\varepsilon_k^{a\dot{a}}=\epsilon_k^a\tilde{\lambda}_k^{\dot{a}}$. A choice of ``parallel'' reference spinors, say $\epsilon_k^a=\lambda_3^a/\br 3k\ke$ as in Ref.\cite{Berkovits:2008ic}, greatly simplifies the computations because it eliminates all bosonic contractions involving the products $\varepsilon_k\varepsilon_l$ and
potentially producing double poles $(z_i-z_j)^{-2}$. In order to discuss this correlator, it is convenient to introduce the following notation for the boundary integrals,  useful not only for $N=6$ but also for arbitrary $N$:
\be
\lsq(i_1i_2)(i_3i_4)\dots(i_{n-1}i_n)\rsq_N =\int_{1}^{\infty}\! dz_4\!\cdots\!\int_{z_{N-1}}^{\infty}\!\! dz_N\,(z_{i_1i_2}z_{i_3i_4}\!\cdots z_{i_{n-1}i_n})^{-1}
\prod_{2\le k<l\le N}|z_{kl}|^{2\alpha'p_kp_l}\label{defint}\ee
where $z_{ij}\equiv z_i-z_j$, $\{i_1,i_2\dots,i_n\}$ take arbitrary values in the set $\{2,3,\dots N\}$,
$z_2=0$ and $z_3=1$. At the face value, for the gauge choice of $\epsilon_k^a=\lambda_3^a/\br 3k\ke$, there are 16 integrals originating from the free-field correlators appearing in Eq.(\ref{cor6}). By using partial fractioning, however, it is possible to express all of them in terms of a 6-element basis. In some bases, the amplitude becomes particularly simple, for instance\footnote{Our notation follows Ref.\cite{dixon}.}
\ba
A_6&=&\frac{\br 12\ke^3}{
\br 34\ke\br 35\ke\br 36\ke}\bigg\{[42][5|2+4|3\ke [61]\lsq(24)(45)(56)\rsq_6+[42][6|2+4|3\ke [51]\lsq(24)(46)(65)\rsq_6\nonumber\\[2mm]
& &\;\quad\qquad\qquad +~ [52][4|2+5|3\ke [61]\lsq(25)(54)(46)\rsq_6+[52][6|2+5|3\ke [41]\lsq(25)(56)(64)\rsq_6\nonumber \\[2mm]
& &\;\quad\qquad\qquad +~ [62][4|2+6|3\ke [51]\lsq(26)(64)(45)\rsq_6+[62][5|2+6|3\ke [41]\lsq(26)(65)(54)\rsq_6\bigg\}\nonumber\\  &&\label{newamp}
\ea
Note the presence of spurious singularities at $\br 35\ke=0$ and at $\br 36\ke=0$ which appear as a consequence of the gauge choice.

The six-gluon MHV amplitude has been given before in Refs.\cite{Stieberger:2006bh,Stieberger:2006te}, where it appears in a rather complicated form. {}For quite a long time, however, we knew \cite{private} that it can be simplified to
\begin{eqnarray}
\nonumber A_6&=&\frac{\langle 12\rangle^3}{\langle 34\rangle\langle 56\rangle}\bigg\{ [12][35][46]\lsq(25)(26)(35)(46)\rsq_6+[15][24][36]\lsq(25)(26)(24)(36)\rsq_6\\[2mm]
& &\quad\qquad~
+ [16][23][45]\lsq(25)(26)(23)(45)\rsq_6-[12]
[36][45]\lsq(25)(26)(36)(45)\rsq_6\nonumber \\[2mm]\nonumber
& &\quad\qquad~ -[15][23][46]\lsq(25)(26)(23)(46)\rsq_6
-[16][24][35]\lsq(25)(26)(24)(35)\rsq_6\bigg\}\\ && \label{oldamp}
\end{eqnarray}
The above expression is free of spurious poles, albeit not as ``symmetric'' as Eq.(\ref{newamp}). By a repeated use of partial fractioning, $(z_{ik}z_{ij})^{-1}=(z_{kj}z_{ik})^{-1}-(z_{kj}z_{ij})^{-1}$,
and partial integrations of the integrals, combined with some spinor algebra, we managed to show that the right hand sides of Eq.(\ref{newamp}) and Eq.(\ref{oldamp}) are indeed equal.

Now we proceed to the general $N$-gluon case.
As in the previous case, a good gauge choice leads to tremendous simplifications. Here, we make a slightly different choice, $\epsilon_k^a=\lambda_2^a/\br 2k\ke$, which ensures $\varepsilon_k\varepsilon_l=0$
for all polarization vectors, in addition to  $\varepsilon_kp_2=\varepsilon_kp_k=0$. Then the bosonic correlation functions  contribute  to the integrands as
\be
\left\langle\varepsilon_{k_1}\partial X(z_{k_1})\varepsilon_{k_2}\partial X(z_{k_2})\dots\prod_{j=1}^N V_j(z_j)\right\rangle=\sum_{i_1\neq 1,2,k_1}\sum_{i_2\neq 1,2,k_2}\!\dots\frac{\varepsilon_{k_1}p_{i_1}}{z_{k_1i_1}}\frac{\varepsilon_{k_2}p_{i_2}}{z_{k_2i_2}}
\dots\!\prod_{2\le k<l\le N}|z_{kl}|^{2\alpha'p_kp_l}\ee
Hence effectively, each $i$ takes $N-3$ possible values and a correlator involving a product of $m$ $\partial X$'s yields $(N-3)^m$ terms. The above correlator includes terms with double poles, $(z_{ij})^{-2}$ which, in bosonic string theory, would give rise to tachyonic singularities. Such terms, however, cancel after including the correlation functions arising from the fermionic parts of the vertices, {\em c.f}.\ Eq.(\ref{vert}). Actually, fermion correlators cancel not only double poles $(z_{ij}z_{ji})^{-1}$ but also all terms involving longer cycles,
$(z_{ij}z_{jk}\cdots z_{mi})^{-1}$. In this way, in $N$-gluon amplitudes, one ends  up with the integrals of the form (\ref{defint}) but without any closed cycles, {\em i.e}.\ all integrals involving closed loops of indices, like $(i_1i_2)(i_2i_3)\dots (i_ki_1)$, are eliminated. There are $(N-2)^{(N-4)}$  integrals remaining.

It seems that the calculations following the proposal of Berkovits and Maldacena yield integrals with one factor of $z_{ij}^{-1}$ less than in the  standard RNS formalism, {\em c.f}.\ Eq.(\ref{newamp}) and Eq.(\ref{oldamp}). {}From now on, we insert the ``missing'' factor of $1=z_{32}=-z_{23}$, so that in both formalisms, all $N$-gluon integrals (\ref{defint})  contain $n/2=N-2$  brackets.

The integrals (\ref{defint}) can be represented by diagrams. To each $z_i$, $i=2,\dots,N$, we associate a point and to each bracket, {\em i.e}.\ to each $(z_{ij})^{-1}$ factor, we associate a link. Since there are no loops of indices, all integrals under consideration can be represented by tree diagrams with $N-2$ links. Note that any point repeating more than two times will produce a branching. As an example, we draw below a typical diagram contributing to $N=8$ amplitudes:
\addtocounter{equation}{1}
\begin{flushleft}
\fcolorbox{white}{white}{
  \begin{picture}(170,74) (163,-43)
    \SetWidth{1.0}
    \SetColor{Black}
    \Vertex(208,-6){2}
    \Vertex(208,26){2}
    \Vertex(208,-38){2}
    \Vertex(248,-6){2}
    \Vertex(288,-6){2}
    \Vertex(328,-6){2}
    \Vertex(168,-6){2}
    \Line[arrow,arrowpos=0.5,arrowlength=5,arrowwidth=2,arrowinset=0.2](168,-6)(208,-6)
    \Line[arrow,arrowpos=0.5,arrowlength=5,arrowwidth=2,arrowinset=0.2](208,26)(208,-6)
    \Line[arrow,arrowpos=0.5,arrowlength=5,arrowwidth=2,arrowinset=0.2](208,-38)(208,-6)
    \Line[arrow,arrowpos=0.5,arrowlength=5,arrowwidth=2,arrowinset=0.2](328,-6)(288,-6)
    \Line[arrow,arrowpos=0.5,arrowlength=0,arrowwidth=0,arrowinset=0.2](288,-6)(208,-6)
     \Text(212,-2)[lb]{$2$}
    \Text(212,-42)[lb]{$3$}
    \Text(165,-2)[lb]{$4$}
    \Text(212,24)[lb]{$5$}
    \Text(246,-2)[lb]{$6$}
    \Text(286,-2)[lb]{$7$}
    \Text(326,-2)[lb]{$8$}
    \Text(346,-12)[lb]{$=[\![(42)(52)(32)(87)(76)(62)]\!]_8\qquad\qquad\qquad\qquad~~~~(\theequation)$}
  \end{picture}
}
\end{flushleft}
We introduced arrows to indicate the ordering of indices inside brackets.
Integrals  associated to tree diagrams will be called  {\em transcendental}, for the reasons explained below.

Some properties of the integrals \req{defint} become more transparent after changing the integration variables according to  $z_4=1/x_1, z_5=1/(x_1x_2),\dots, z_N=1/(x_1x_2\cdots x_{N-3})$ \cite{Stieberger:2006te}. We obtain:
\be\label{euler}
\lsq(i_1i_2)(i_3i_4)\dots(i_{n-1}i_n)\rsq_N =\lf(\prod_{i=1}^{N-3} \int^1_0 d x_i\ri)
\prod_{a=1}^{N-3} x_a^{\alpha' s_{23\ldots a+2}+n_a}\
\prod_{b=a}^{N-3} \lf( 1 -\prod_{j=a}^b x_j \right)^{\alpha' s_{a+2,b+3}+n_{a,b}},
\ee
depending on  the kinematic invariants
$s_{i,j}\equiv s_{ij}=(p_i+p_j)^2$ and $s_{i_1\ldots i_l}=(p_{i_1}+\ldots+p_{i_l})^2$. The integers $n_a$ and $n_{ab}$ are determined by
\bea
n_a&=&1+a-N-\sum_{j=3+a}^N\sum_{i=2}^{j-1} \tilde n_{ij}\ ,\nonumber\\
n_{a,b}&=&\tilde n_{a+2,b+3}\ ,\ \ \ a=1,\ldots,N-3,\ b=a,\ldots,N-3\ ,
\label{integers}
\eea
where $\tilde n_{ij}=-1$ for each link $(ij)$, otherwise $\tilde n_{ij}=0$.
The integrals (\ref{euler}) represent generalized Euler integrals, which integrate to multiple Gaussian hypergeometric functions~\cite{Oprisa:2005wu}.
They depend parametrically on the kinematic invariants, which are constrained by the momentum conservation law and the mass--shell condition $p_i^2=0$.

The representation (\ref{euler}) is particularly suitable for studying the low energy $\alpha'\to 0$ limit of the integrals. {}For $N$ gluons, the leading  terms ${\cal O}[(\alpha')^{3-N}]$ yield the Yang-Mills limit of the amplitude, with the kinematic singularities associated to massless gauge bosons propagating in intermediate channels.  We call the integrals associated to tree diagrams``transcendental'' because their low-energy expansions are rather special: the powers of $\alpha'$ are always accompanied by zeta functions of integer arguments, with a fixed ``degree of transcendentality.''
The degree of transcendentality (transcendentality level \cite{Fleischer:1998nc}) of $\pi$ is defined to be $1$, while that of
$\zeta(n)$ is $n$, and  for multiple zeta values $\zeta(n_1,\ldots,n_r)$ it is  $\sum_{l=2}^rn_l$.
The degree of transcendentality  for a product is defined to be the
sum of the degrees of each factor.
The expansions have the  form:
\be\label{Euler}
\lsq(i_1i_2)(i_3i_4)\dots(i_{n-1}i_n)\rsq_N =(\ap)^{3-N}\ p_{3-N}+(\ap)^{5-N}\ p_{5-N}\ \zeta(2)\ +(\ap)^{6-N}\ p_{6-N}\
\zeta(3)\ +\ldots\ .
\ee
where $p_l$  are
degree $l$ homogenous rational functions of the kinematic invariants $s_{i\ldots j}$.
In \req{Euler} at each order $5-N+m$ in $\ap$ only a set of Riemann zeta functions (or multiple zeta values) appears, with the degree of transcendentality equal $m+2$.

We claim that any  integral \req{defint} and \req{euler} can be written as a linear combination of
$(N-3)!$ basis integrals, weighted by rational functions of the kinematic invariants. At this time, we are not able to prove this claim in any other way than by tedious induction. We confirm the conclusions of Ref.\cite{Stieberger:2006te,Stieberger:2007jv,Mafra:2011nv,Mafra:2011nw}.

In order to cast the $N$--gluon amplitude in a simplest possible form, the $(N-2)^{(N-4)}$ transcendental integrals, remaining after combining bosonic and fermionic contractions, should be expressed in a suitable basis. To that end, we introduce the ``chain'' basis $\{\mathscr{C}^N_{\sigma}\}$   labeled by $(N-3)!$ permutations $\sigma$ of the set $\{4,5,\dots,N\}$:
\addtocounter{equation}{1}
\begin{flushleft}
\fcolorbox{white}{white}{
  \begin{picture}(267,32) (126,-59)
    \SetWidth{1.0}
    \SetColor{Black}
    \Vertex(208,-48){2}
    \Vertex(377,-48){2}
    \Vertex(248,-48){2}
    \Vertex(328,-48){2}
    \Vertex(168,-48){2}
    \Text(205,-42)[lb]{$3$}
    \Text(165,-42)[lb]{$2$}
    \Text(245,-44)[lb]{$4_{\sigma}$}
    \Text(372,-44)[lb]{$N_{\sigma}$}
    \Text(305,-45)[lb]{$ (N-1)_{\sigma}$}
    \Line(328,-48)(377,-48)
    \Line[arrow,arrowpos=0.25,arrowlength=5,arrowwidth=2,arrowinset=0.2](168,-48)(248,-48)
    \Text(122,-52)[lb]{$\mathscr{C}^N_{\sigma}~=$}
     \Text(394,-53)[lb]{$=~ [\![(23)(34_{\sigma})\dots((N-1)_{\sigma}N_{\sigma})]\!]_N~(\theequation)$}
    \Line[dash,dashsize=2](240,-48)(336,-48)
  \end{picture}
}
\end{flushleft}
where $i_{\sigma}\equiv\sigma(i)$. Expressing $(N-2)^{(N-4)}$ functions in this particular chain basis is a purely algebraic operation involving partial fractioning only; no partial integrations are necessary to accomplish it. After this step, followed by some spinor algebra, we obtain
\be
A_N~=~\frac{\br 12\ke^4}{\br 12\ke
\br 23\ke\br 31\ke}\bigg(\prod_{k=4}^N\br 2k\ke\bigg)^{-1}\sum_{\sigma}\mathscr{C}^N_{\sigma}\br 2|3|4_{\sigma}]\,\br 2|3+4_{\sigma}|5_{\sigma}]\cdots\br 2|3+4_{\sigma}+\dots+(N-1)_{\sigma}|N_{\sigma}]\ .
\label{namp}
\ee

 The amplitude (\ref{namp}) contains spurious poles at $\br 2k\ke=0$ due to the gauge choice $\varepsilon_k p_2=0$.
All the functions $\mathscr{C}^N_{\sigma}$ furnish the transcendental power series expansion \req{Euler} with $p_{3-N}\neq 0$.
  Actually, depending on the gauge choice and the choice of basis, the amplitude can be rewritten in various ways, with different spurious singularities. One can ask  if there exist a basis in which the amplitude is manifestly free of such singularities. Eq.(\ref{oldamp}) shows that it does exist
 for $N=6$. We addressed this question in the case of $N=7$. Indeed,  spurious singularities can be eliminated by manipulating Eq.(\ref{namp}), leaving physical poles only, say at $\br 34\ke=0$, $\br 56\ke=0$,
 $\br 67\ke=0$. Such manipulations, however, in particular the partial integrations, destroy the symmetry of chain basis and the resultant expression is not as simple as Eq.(\ref{namp}).

An important new insight into the structure of $N$-gluon tree level Yang-Mills amplitudes has been  recently gained by rewriting them in terms of the ``position'' coordinates dual to momentum variables. These coordinates are defined implicitly by $p_k=x_k-x_{k-1}$, with the momentum conservation expressed by $x_0=x_N$. Then the amplitudes exhibit a ``dual'' superconformal symmetry \cite{Drummond:2008vq}. In 2009, Andrew Hodges \cite{Hodges:2009hk} introduced ``momentum'' twistors associated to the dual superconformal group. Tree-level Yang-Mills amplitudes become manifestly covariant when written in terms of momentum twistor variables. No such symmetry is expected to hold for full-fledged superstring amplitudes because dilatational symmetry is manifestly violated by the dependence on $\xa'$. Nevertheless
 we can try to express Eq.(\ref{namp}) in terms of momentum twistors, in order to compare Yang-Mills with the full-fledged superstring amplitudes.

The momentum-twistors are defined as $Z_k=(\lambda_k,\tilde{\mu}_k{=}\lambda_kx_k)$ and the dual momentum-twistors as $W_k=({\mu}_k{=}x_k\tilde{\lambda}_k,\tilde{\lambda}_k)$.
With the choice of $x_1=0$ as the origin of the dual coordinate space, $\mu_1=\tilde{\mu}_1=\mu_2=\tilde{\mu}_2=0$ and $x_n=\sum_{k=2}^np_k$. Now the $N$-gluon amplitude (\ref{namp}) can be rewritten as
\be
A_N~=~\frac{\br 12\ke^4}{\br 12\ke
\br 23\ke\br 31\ke}\bigg(\prod_{k=4}^N\br 2k\ke\bigg)^{-1}\sum_{\sigma}\mathscr{C}^N_{\sigma} (Z_2W_4^{\sigma})(Z_2W_5^{\sigma})\cdots (Z_2W_N^{\sigma})\ .
\label{ntwist}\ee

To summarize, we derived a simple formula, given in Eqs.(\ref{namp}) and (\ref{ntwist}), for the superstring MHV amplitude at the disk level.
It is possible that the twistor-dependence of kinematic factors is purely coincidental. On the other hand, it may point to a deeper role of momentum twistors in superstring theory.\\[5mm]
\textbf{Acknowledgements}\\[5mm]  St.St.\ is grateful to Carlos Mafra and Oliver Schlotterer for valuable discussions. T.R.T.\ thanks Jacob Bourjaily and James Drummond
for useful discussions and correspondence.  Both authors are grateful to the Theory Division of CERN for hospitality and financial support during various stages of this work.
This material is based in part upon work supported by the National Science Foundation under Grant No.\ PHY-0757959.  Any
opinions, findings, and conclusions or recommendations expressed in
this material are those of the authors and do not necessarily reflect
the views of the National Science Foundation.


\begin{thebibliography}{99}
\bibitem{Parke:1986gb}
  S.J.~Parke and T.R.~Taylor,
  ``Amplitude for N-Gluon Scattering,''
  Phys.\ Rev.\ Lett.\  {\bf 56} (1986) 2459.
\bibitem{Witten:2003nn}
  E.~Witten,
  ``Perturbative gauge theory as a string theory in twistor space,''
  Commun.\ Math.\ Phys.\  {\bf 252} (2004) 189
  [arXiv:hep-th/0312171].
\bibitem{Oprisa:2005wu}
  D.~Oprisa and S.~Stieberger,
  ``Six gluon open superstring disk amplitude, multiple hypergeometric series and Euler-Zagier sums,''
  hep-th/0509042.
\bibitem{Stieberger:2006bh}
  S.~Stieberger and T.R.~Taylor,
  ``Amplitude for N-gluon superstring scattering,''
  Phys.\ Rev.\ Lett.\  {\bf 97} (2006) 211601
  [arXiv:hep-th/0607184].
\bibitem{Stieberger:2006te}
  S.~Stieberger and T.R.~Taylor,
  ``Multi-gluon scattering in open superstring theory,''
  Phys.\ Rev.\  D {\bf 74} (2006) 126007
  [arXiv:hep-th/0609175].
\bibitem{Stieberger:2007jv}
  S.~Stieberger and T.R.~Taylor,
  ``Supersymmetry Relations and MHV Amplitudes in Superstring Theory,''
  Nucl.\ Phys.\  B {\bf 793} (2008) 83
  [arXiv:0708.0574 [hep-th]].
  \bibitem{Stieberger:2007am}
  S.~Stieberger and T.R.~Taylor,
  ``Complete Six-Gluon Disk Amplitude in Superstring Theory,''
  Nucl.\ Phys.\ B {\bf 801}, 128 (2008)
  [arXiv:0711.4354 [hep-th]].
\bibitem{Mafra:2011nv}
  C.R.~Mafra, O.~Schlotterer and S.~Stieberger,
  ``Complete N-Point Superstring Disk Amplitude I. Pure Spinor Computation,''
  arXiv:1106.2645 [hep-th].
\bibitem{Mafra:2011nw}
  C.R.~Mafra, O.~Schlotterer and S.~Stieberger,
  ``Complete N-Point Superstring Disk Amplitude II. Amplitude and
  Hypergeometric Function Structure,''
  arXiv:1106.2646 [hep-th].
\bibitem{Berkovits:2008ic}
  N.~Berkovits and J.~Maldacena,
  ``Fermionic T-Duality, Dual Superconformal Symmetry, and the Amplitude/Wilson
  Loop Connection,''
  JHEP {\bf 0809} (2008) 062
  [arXiv:0807.3196 [hep-th]].
\bibitem{dixon}
  L.~J.~Dixon,
  ``Calculating scattering amplitudes efficiently,''
  arXiv:hep-ph/9601359.
\bibitem{private}S.~Stieberger and T.R.~Taylor, unpublished (2009).
\bibitem{Fleischer:1998nc}
  J.~Fleischer, A.V.~Kotikov and O.L.~Veretin,
  ``Applications of the large mass expansion,''
  Acta Phys.\ Polon.\ B {\bf 29} (1998) 2611
  [hep-ph/9808243];
``Analytic two loop results for selfenergy type and vertex type diagrams with one nonzero mass,''
  Nucl.\ Phys.\ B {\bf 547}, 343 (1999)
  [hep-ph/9808242];\\
A.V.~Kotikov, L.N.~Lipatov, A.I.~Onishchenko and V.N.~Velizhanin,
``Three loop universal anomalous dimension of the Wilson operators in N=4 SUSY Yang-Mills model,''
  Phys.\ Lett.\ B {\bf 595}, 521 (2004)
  [Erratum-ibid.\ B {\bf 632}, 754 (2006)]
  [hep-th/0404092].
  \bibitem{Drummond:2008vq}
  J.M.~Drummond, J.~Henn, G.P.~Korchemsky and E.~Sokatchev,
  ``Dual superconformal symmetry of scattering amplitudes in N=4
  super-Yang-Mills theory,''
  Nucl.\ Phys.\  B {\bf 828} (2010) 317
  [arXiv:0807.1095 [hep-th]].
\bibitem{Hodges:2009hk}
  A.~Hodges,
  ``Eliminating spurious poles from gauge-theoretic amplitudes,''
  arXiv:0905.1473 [hep-th].


\end{thebibliography}
\end{document}